# Disordered insulator in an optical lattice

M. Pasienski[1], D. McKay[1], M. White[1], and B. DeMarco[1]

[1] *Physics Department, University of Illinois, 1110 W Green St, Urbana, IL 61801*

**Disorder can profoundly affect the transport properties of a wide range of quantum materials. For example, the superfluid transition temperature for helium is strongly modified when it is adsorbed in porous media[1], and understanding how disorder affects the emergence of high-temperature superconductivity is a problem with critical implications for both practical applications and fundamental condensed matter physics[2]. Presently, there is significant disagreement regarding the effect of disorder on transport in the disordered Bose-Hubbard (DBH) model, which is the paradigm used to theoretically study disorder in strongly correlated bosonic systems[3,4]. We experimentally realize the DBH model by using optical speckle to introduce precisely known, controllable, and fine-grained disorder to an optical lattice[5]. Here, by measuring the dissipation strength for transport[6], we discover a disorder-induced a SF-to-insulator (IN) transition in this system, but we find no evidence for an IN-to-SF transition, in conflict with certain theoretical predictions. Emergence of the IN at disorder strengths several hundred times the tunnelling energy agrees with a predicted SF–Bose glass (BG) transition from recent quantum Monte Carlo (QMC) work[7]. Both the SF–IN transition and correlated changes in the atomic quasimomentum distribution—which verify a simple model for the interplay of disorder and interactions in this system—are phenomena new to the unit filling regime explored in this work, compared with the high filling limit probed previously[5]. We find that increasing disorder strength generically leads to greater dissipation in the regime of mixed SF and Mott-insulator (MI) phases, excluding predictions of a disorder-induced, or "re-entrant," SF (RSF). While the

**absence of an RSF may be explained by the effect of finite temperature, we strongly constrain theories by measuring bounds on the entropy per particle in the disordered lattice.**

A wide range of sophisticated theoretical approaches applied to determine the ground state phase diagram of the DBH model have reached inconsistent conclusions (see Refs. [3,4] and [8] for an overview and [7,9,10,11,12] for recent results). Precisely how insulating phases arise as microscopic parameters—such as the disorder strength, interaction energy, tunnelling energy, and chemical potential—are varied is presently a subject of intense debate, particularly for three-dimensional systems. Resolving these issues by comparison to measurements on solids is complicated by incomplete knowledge of the physical disorder and the influence of other degrees of freedom, such as lattice distortions and long-range electronic interactions. In particular, the BG phase—a compressible, gapless IN first discussed in Ref. [13]—has yet to be unambiguously observed in an experiment.

Ultra-cold atom gases confined in disordered potentials have recently emerged as an ideal system for exploring these fundamental questions[3,4,14]. Disorder can be controllably introduced using an optical speckle field[15,16,17,18]; the disorder strength can be continuously varied by changing the optical intensity, and precise characterization of the disordering potential is possible using high-resolution microscopy. Because the interactions between atoms can also be manipulated, the effects of disorder can be explored both in the non-interacting regime and in the strongly correlated limit. For example, Anderson localization in a disordered potential generated by speckle[19] and in a quasi-crystalline system[20] has been observed, and we have realized the DBH model in the strongly interacting regime by combining optical speckle with a three-dimensional optical lattice[5].

In our experiment, $^{87}$Rb Bose-Einstein condensates are confined in a disordered optical lattice created by three pairs of 812 nm laser beams and a 532 nm optical speckle field, as described in Ref. [5]. Figure 1 shows the relative geometry of the lattice beams (red lines) and disordering potential (green arrow) projected onto the plane used to image the atom gas. We control the ratio of interaction energy $U$ to tunnelling energy $t$ for the atoms by tuning the lattice potential depth, which is characterized by a dimensionless parameter $s$ (the lattice potential depth is $sE_R$ along each lattice direction, where $E_R$ is the atomic recoil energy at 812 nm). The speckle potential is cylindrically symmetric with characteristic speckle sizes 570 nm and 3 μm along the transverse and longitudinal directions to propagation; sample slices through the measured speckle intensity profile are shown in green in Fig. 1. Because the disorder in our experiment is fine grained (i.e., smaller than two 406 nm lattice spacings along any lattice direction), the speckle potential leads to a distribution of Hubbard site occupation, tunnelling, and interaction energies (see Ref. [5]). We characterize the strength of the disorder $\Delta$ by the average potential shift from the speckle field, which is equivalent to the standard deviation of the site occupation energy distribution. In contrast to our previous work in the high filling limit[5], we adjust the atom number so that approximately one particle occupies each site in the centre of the lattice for all of the data reported here.

The effect of disorder on the transport properties of the gas is shown in Fig. 3. Data are shown for the COM velocity of the gas along the transverse ($v_T$) and longitudinal ($v_L$) speckle directions immediately after an applied impulse for $\Delta=0$, 0.75, and 3 $E_R$ and for a range of lattice depths spanning the SF and MI regimes in a clean lattice ($s$=6–19). The technique we employ here is a modification of that used in Ref. [6], in which the dissipation for transport $\gamma$ is probed by measuring the COM motion of the gas in the parabolic confining potential after applying a spatially uniform impulse. These data are taken in the regime in which the dominant dissipation mechanism in a clean lattice arises from quantum phase slips. We work primarily in the limit of an impulse that is rapid compared with the harmonic motion but not the dissipation rate. In this regime ($\gamma \gg 1/t \gg \omega$), $v = \frac{Ft}{m^*}e^{-\gamma t}$, where $F$ is the force during the impulse of duration $t$, $\omega$ is the oscillator frequency for motion in the parabolic potential, and $m^*$ is the effective mass (which depends on $s$). Since we fix the impulse $Ft$ for the data in Fig. 3, the emergence of an IN at fixed $s$ is characterized by $\gamma \to \infty$ and $v = 0$, which is denoted by the dashed line.

In the clean lattice, the velocity approaches the expectation for an IN as the lattice depth is increased into the regime for which atoms in the unit-filling MI state exist in the centre of the lattice. This suppression of the COM velocity for increasing $s$ is caused by an increase in effective mass (characterized by the prediction for $\gamma = 0$, which is shown by the solid black line) and by an enhancement in the quantum phase slip rate (which scales as $\gamma \propto e^{\sqrt{U/t}}$). The atom gas does not completely transform into an IN when atoms emerge in the MI state because a SF shell exists at the periphery of the gas at low temperature[21]; the signature of this SF is the finite condensate fraction $N_0/N$ (defined as in Ref. [5]) observed for $s > 13.3$.

Although disorder weakly affects transport in the SF regime for $\Delta = 0.75 E_R$, increasing the disorder strength to $\Delta = 3 E_R$ transforms the SF state to an IN at approximately $s = 12$, corresponding to $U/t = 25$ and $\Delta/t = 250$. This transition agrees within our systematic uncertainty (see Methods) in $U/t$ and $t$ with the prediction for a SF–BG phase boundary from recent QMC work[7]. Within the systematic uncertainty in determining zero velocity (see Methods), the SF-to-IN transition occurs at the same $s$ for the longitudinal and transverse directions. The region of $U/t$ and effective chemical potential $\mu$ that we determine as insulating is shown superimposed in blue on the phase diagram for the clean system in the middle inset to Fig. 1;. Because of the confining harmonic potential, we sample a range of fillings and corresponding $\mu$—we assume that the gas must be globally insulating for $v$ to vanish after the impulse. The region we determine as insulating is consistent with an early, qualitative prediction for the DBH phase diagram[22]. The plot of $N_0/N$ for $\Delta = 0$, 0.75, and 3 $E_R$ in Fig. 3 shows that the emergence of the disordered insulating state coincides with the obliteration of the condensate by disorder. To assess the impact of finite temperature, we measure the transport properties for atoms in the clean lattice ($\Delta = 0\, E_R$) at sufficiently high temperature to match the condensate fraction observed at $\Delta = 3\, E_R$. Even though at this temperature the entropy per particle in the clean lattice is higher than for $\Delta = 3\, E_R$ (see the discussion that follows and Fig. 4), the gas is not insulating, therefore indicating that the observed SF-to-IN transition is not caused exclusively by heating introduced by the speckle field.

Some hints about the mechanism behind the extraordinary robustness of superfluidity in this system are revealed in representative images (shown in Fig. 2) of the atomic quasimomentum distribution taken after bandmapping[23] and time-of-flight (TOF). The transition to an IN at $s$=12 for $\Delta = 3$ is marked by dramatic, but not unexpected, behaviour: the condensate is destroyed, giving rise to a broad quasimomentum distribution (Fig. 2d). The quasimomentum distribution in Fig. 2d closely resembles the distribution deep in the MI regime, in which the atoms have equal quantum amplitude at all quasimomentum and are therefore localized, as anticipated for an IN.

More informative is the effect of disorder observed at $s$=6, well into the SF regime of the clean BH model. The non-condensate (NC) populated by applying strong disorder ($\Delta = 3E_R$) reflects the symmetry of the speckle field, with a diameter approximately twice as large along the transverse direction (Fig 2b). This behaviour directly confirms a conjecture that atoms can localize to screen the disorder through interactions[8,24,25]. A random distribution of site energies—indicated by black lines in

Fig. 2e—would result in high kinetic energy states localized into the deepest lattice wells for non-interacting particles. The strong repulsive interactions present in this many-particle system, however, prevent multiple occupation of these localized states. Instead, only a fraction of the atoms localize into the deepest wells, leading to a corresponding reduction of condensate fraction. The remaining atoms exist in a delocalized, lower energy SF state because they experience a more uniform effective potential (red lines in Fig. 2e) arising from their interactions with the localized atoms. The asymmetry of the NC at $s$=6 is a direct result of the difference in speckle correlation length scales along the transverse and longitudinal directions: in contrast to the transverse direction for which energy shifts are uncorrelated from site-to-site, the atoms can localize to a several-site-region to screen the disorder along the longitudinal direction.

In contrast to many theoretical predictions for the DBH model that exhibit RSF generated by introducing disorder to a MI[9,10,11,12], we find that increasing disorder generically leads to larger dissipation. To directly test for an RSF, we measure transport as $\Delta$ is varied at $s$=14 ($U/t = 44$), for which approximately half of the atoms start in the MI phase in the clean lattice at zero temperature (according to the local density approximation and site-decoupled mean-field theory[26]). The data shown in the inset to Fig. 1 are typical for all lattice depths: increasing $\Delta$ leads to greater dissipation (and corresponding decrease in $|v|$) for motion along either the longitudinal or transverse directions. This behaviour is in conflict with, for example, Ref. [9], which predicts that the gas should wholly convert into a SF at $U/t \approx 40$ for this range of disorder, and hence that the dissipation strength should decrease.

The conspicuous absence of the RSF in this system may be the result of several factors, such as finite system size, finite temperature, and the microscopic disorder parameters for our system, none of which (to our knowledge) have been included in a complete theoretical treatment. While finite size effects have been found to impact the MI phase boundaries[27], the RSF is present in numerical simulations with lattices smaller than we sample here[11]. Furthermore, this phase has been discovered in theoretical treatments for a variety of disorder distributions. Therefore, we feel that the most probable explanation for this discrepancy is finite temperature, since there is some evidence that the RSF has an exceptionally low critical temperature[7]. Unfortunately, we cannot measure temperature precisely because there is not a quantitative, verified theory that connects measureable quantities to temperature.

We have therefore measured bounds on entropy per particle $S/N$ in order to constrain future finite temperature theoretical treatments. To determine the limits shown in Fig. 4, we measure condensate fraction in the purely parabolic potential before transferring the atoms into the disordered lattice and after slowly turning off the lattice. The second law of thermodynamics ensures that this procedure provides lower and upper bounds on entropy. The entropy per particle is determined using $S/N = 3.6(1 - N_0/N)$, which is exact for a non-interacting gas and approximately correct for an interacting, harmonically trapped gas for the range of $N_0/N$ measured here[28]. For the re-entrant SF regime, we find that $S/N < 2k_B$, where $k_B$ is Boltzmann's constant.

In conclusion, while the agreement between our measurements and recent QMC work indicate that the disordered insulator we have discovered is a BG, we cannot currently experimentally distinguish between MI, BG, and other insulating phases. Measurements of the excitation spectrum and compressibility[29] of the insulating state will be necessary before it can be definitively identified as a BG. Our measurements will ultimately act as a benchmark for work in progress by other groups extending stochastic mean-field theory, replica theory, and QMC algorithms to include finite temperature and our disorder parameters.

**Methods**

We create $^{87}$Rb BECs using a hybrid magneto-optical trap that is generated using a magnetic quadrupole field and a 1064 nm focused laser beam (see Ref. $^{23}$ for details). The average number of atoms is $(12\pm4)\times10^3$, corresponding to 1.4 and 1 particles per site in the centre of the clean lattice for $s=6$ and 14 according to site-decoupled mean field theory. The ratio $U/t$ is calculated from $s$ using a band structure calculation and the known atomic interaction parameters. Our typical systematic uncertainty in $s$ is 6%, which corresponds to 25% in $U/t$ and 12% in $t$ at $s=12$, for example.

An impulse is applied along the transverse direction via translating the 1064 nm laser beam by 7 $\mu m$ using a trapezoidal ramp that lasts for a total of 3 ms. A uniform magnetic field is applied for 1 ms to create a longitudinal impulse. The magnitude and timing of the magnetic field pulse and the laser beam displacement were adjusted to produce approximately equal impulses along the two directions. We turn off the disordered lattice after the impulse using a 200 $\mu s$ linear ramp of the lattice and speckle light. The impulse does not significantly affect the condensate fraction measured after TOF.

The systematic uncertainty in determining zero velocity arises from using a technique to determine this parameter that is unbiased but can result in positive $v$, which is unphysical given the fixed impulse direction. We set zero velocity by measuring the COM position $x_0$ of the NC without an applied impulse. We correct for a systematic shift in $x_0$ that depends on $\Delta$ arising from a slight misalignment of the speckle field with respect to the centre of the lattice. We also observe a weak systematic dependence of $x_0$ on $s$. The $\pm0.15$ $mm/s$ systematic uncertainty shown in Fig. 3 is equivalent to the spread in $x_0$ for all $s$ and $\Delta$ used in this work. The measured velocity of the condensate and NC components after the impulse is determined by the displacement from $x_0$ following 20 ms TOF. The COM velocity is defined as $v = (v_0 N_0 / N_{NC} + v_{NC}) / (1 + N_0 / N_{NC})$, where $v_0$ and $v_{NC}$ are the condensate and NC velocities, and $N_0$ and $N_{NC}$ are the numbers of condensate and NC atoms.

This work was supported by the DARPA OLE program (ARO award W911NF-08-1-0021), the Sloan Foundation, and the National Science Foundation (award 0448354). D. McKay acknowledges support from NSERC.

Correspondence and requests for materials should be addressed to B. DeMarco (bdemarco@illinois.edu)

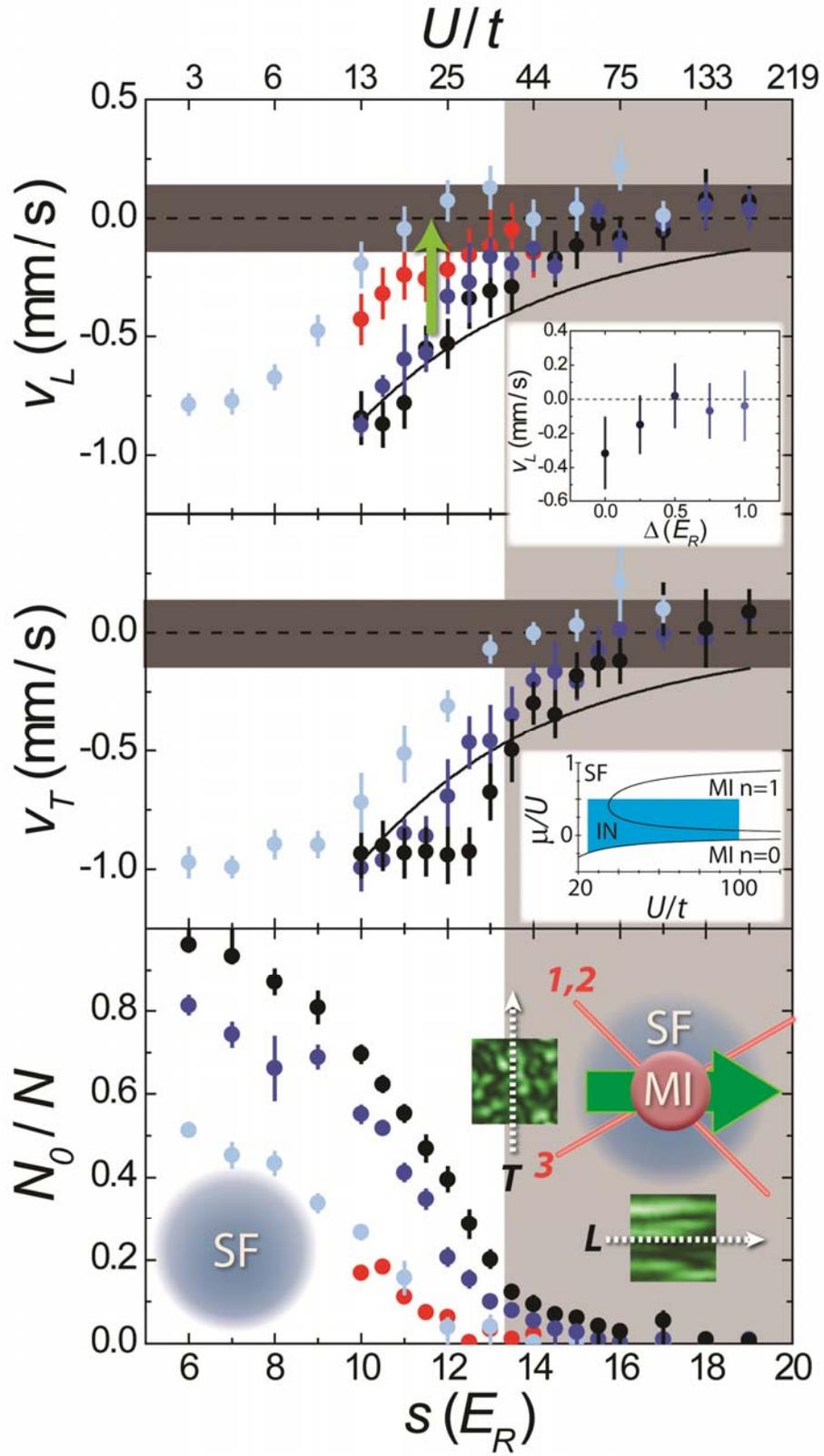

**Figure 1. Effect of disorder on transport.** The overall COM velocity *v* of the gas after an impulse is applied is plotted as the lattice depth *s* is varied for three disorder strengths: $\Delta = 0$ (black), 0.75 (dark blue), and 3.0 $E_R$ (light blue); data are shown for the longitudinal (top) and transverse (middle) directions. Data are also shown for high temperature transport in a clean lattice (red). The top inset shows how disorder affects $v_L$ for $s = 14$; the colour scale for the points follows the main figure for reference. Within the region marked in light grey, a spherical core of unit filling MI exists in the centre of the clean lattice (according to 3D mean field theory). The dark grey band indicates the systematic uncertainty in determining zero velocity (see Methods). The error bars represent the statistical uncertainty in the (typically) seven measurements averaged for each point. The error bars for $v$ also include the statistical uncertainty in determining zero velocity, and the error bars for $N_0 / N$ include a systematic error that reflects our inability to measure condensate fraction above 95% and below 5%.

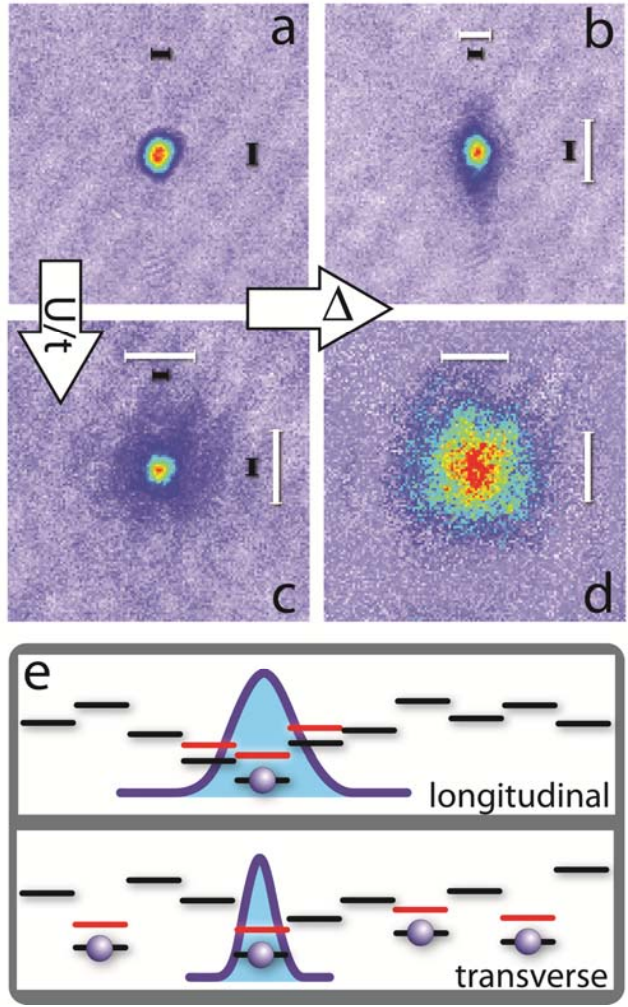

**Figure 2. Effect of disorder on atomic quasimomentum distribution. a–d,** Images taken after bandmapping from the disordered lattice and TOF are shown in false colour for **a,** $s$=6, $\Delta = 0$; **b,** $s$=6, $\Delta = 3E_R$ ; **c,** $s$=12, $\Delta = 0$; and **d,** $s$=12, $\Delta = 3E_R$. The TOF is 25 ms for $s = 6$ and 15 ms for $s = 12$; the field-of-view for each image is 0.6 mm. The images are fit to two-component Gaussian distributions to determine condensate fraction and the sizes and locations of the condensate and NC. The black and white bars correspond to twice the fitted r.m.s. radius for the condensate and NC components. **e,** A simple model—suggested by the behaviour in **b**—that may be used to understand the effect of disorder on the quasimomentum distribution. Atoms (blue spheres) localize to regions (blue curves) to screen the disordered potential and create a more uniform effective potential.

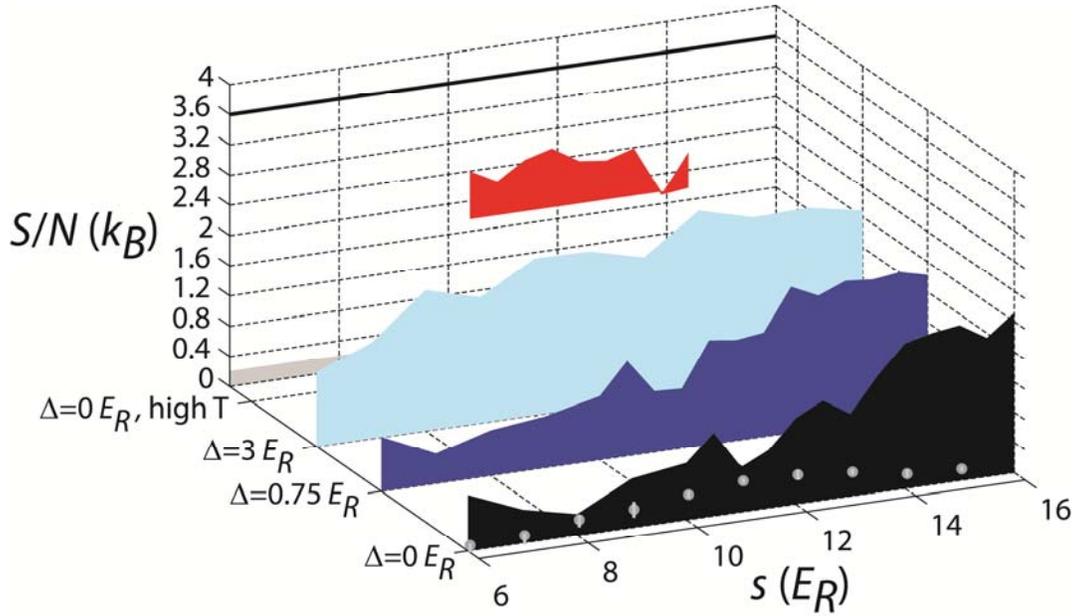

**Figure 3. Bounds on entropy per particle in the disordered lattice.** The entropy per particle $S/N$ is determined to lie within the coloured regions for $\Delta = 0$ (black), 0.75 (dark blue), 3.0 $E_R$ (light blue), and $\Delta = 0$ at high temperature (red). The $S/N$ corresponding to the critical temperature for condensation in a weakly interacting, parabolically confined system is indicated by the solid black line. The systematic uncertainty in the lower bound related to our inability to resolve high $N_0/N$ is shown by the light grey band. There is typically both a 0.2 $k_B$ systematic and 0.2 $k_B$ statistical uncertainty in the upper bound. The minor, disorder-induced heating implied by the increase in the upper bound for $\Delta = 3E_R$ between $s = 8-11$ may be an artefact of the measurement technique. For comparison, we find that the $S/N$ determined from bandmapped $N_0/N$ measured in the clean lattice (light gray points) is close to the lower bound for all $s$, even though the upper bound increases significantly for $s > 12$. We calculate $S/N$ for these points using the local density approximation and site-decoupled mean field theory to self-consistently solve for the temperature and chemical potential.